# A High Performance Memory Database for Web Application Caches

Ivan Voras, Danko Basch and Mario Žagar

*Abstract*—This paper presents the architecture and characteristics of a memory database intended to be used as a cache engine for web applications. Primary goals of this database are speed and efficiency while running on SMP systems with several CPU cores (four and more). A secondary goal is the support for simple metadata structures associated with cached data that can aid in efficient use of the cache. Due to these goals, some data structures and algorithms normally associated with this field of computing needed to be adapted to the new environment.

## I. Introduction

We observe that, as the number of web applications created in scripting languages and rapid prototyping frameworks continues to grow, the importance of off-application cache engines is rapidly increasing [1]. Our own experiences in building high-performance web applications has yielded an outline of a specification for a cache engine that would be best suited for dynamic web applications written in an inherently stateless environment like the PHP language and framework[1]. Finding no Open Source solutions that would satisfy this specification, we have created our own.

## II. Specification

The basic form for an addressable database is a store of key-value pairs (a dictionary), where both the key and the value are more or less opaque binary strings. The keys are treated as addresses by which the values are stored and accessed. Because of the simplicity of this model, it can be implemented efficiently, and it's often used for fast cache databases. The first point of our specification is that *the cache daemon will implement key-value storage*.

However, key-value databases can be limiting and inflexible. In our experience one of the most valuable features a cache database could have is the ability to group cached keys by some criterion, so multiple keys belonging to the same group can be fetched and expired together, saving communication round-trips and removing grouping logic from the application. To make the cache database more versatile, *the cache daemon will implement simple metadata for cached records in the form of typed numeric tags*.

Today's pervasiveness of multi-core CPUs and servers with multiple CPU sockets has results in a significantly different environment than what was common when single-core single-CPU computers were dominant. Algorithms and structures that were efficient in the old environment are sometimes suboptimal in the new. Thus, *the cache daemon will be optimized in its architecture and algorithms for multi-processor servers with symmetric multi-processing (SMP)*.

Finally, because of the many operating systems and environments available today, *the cache daemon will be written according to the POSIX specification and usable on multiple platforms*.

### A. Rationale and discussion

Cache databases should be fast. The primary purpose of this project was to create a cache database for use in web applications written in relatively slow languages and frameworks such as PHP, Python and Ruby. The common way of using cache databases in web applications is for storing results of complex calculations, both internal (such as generating HTML content) and external (such as from SQL databases). The usage of dedicated cache databases pays off if the cost of storing (and especially retrieving) data to (and from) the cache database is lower than the cost of performing the operation that generates the cached data. This cost might be in IO and memory allocation but we observe that the more likely cost is in CPU load. Caching often-generated data instead of generating it repeatedly (which is a common case in web applications, where the same content is presented to a large number of users) can dramatically improve the application's performance.

Though they are efficient, we have observed that pure key-value cache databases face a problem when the application needs to atomically retrieve or expire multiple records at once. While this can be solved by keeping track of groups of records in the application or (to a lesser extent) folding group qualifiers into key names, we have observed that the added complexity of bookkeeping this information in a slow language (e.g. PHP) distracts from the simplicity of the cache

---

[1]Because of the stateless nature of the HTTP, most languages and frameworks widely used to build web applications are stateless, with various workarounds like HTTP cookies and server-side sessions that are more-or-less integrated into frameworks.

This work is supported in part by the Croatian Ministry of Science, Education and Sports, under the research project "Software Engineering in Ubiquitous Computing".

Ivan Voras, Danko Basch and Mario Žagar are with Faculty of Electrical Engineering and Computing, University of Zagreb, Croatia. (e-mail: {ivan.voras, danko.basch, mario.zagar}@fer.hr).

engine and can even lead to slowdowns. Thus we added the requirement for metadata in the form of typed numeric "tags" which can be queried. Adding tags to key-value pairs would enable the application to off-load group operations to the cache daemon where the data is stored.

With the recent trend of increasing the number of CPU cores in computers, especially with multi-core CPUs [10], we decided the cache daemon must be designed from the start to make use of multiprocessing capabilities. The project is to explore efficient algorithms that should be employed to achieve the best performance on servers with 4 and more CPU cores. Due to the read-mostly nature of its purpose, the cache database should never allow readers (clients that only fetch data) to block other readers.

Multi-platform usability is non-critical, it is an added bonus that will make the result of the project usable for many more users.

### III. IMPLEMENTATION OVERVIEW

We have implemented the cache daemon in the C programming language, using only the standard or widespread library functions to increase its portability on POSIX-like operating systems. POSIX Threads (*pthreads*) were used to achieve concurrency on multi-processor hardware.

The deamon can be roughly divided into three modules: network interface, worker threads and data storage. The following sections will describe the implementation details of each of the modules.

#### A. Network interface

The network interface module is responsible for accepting and processing commands from clients. It uses the standard BSD "sockets" API in non-blocking mode. By default, the daemon creates both a "Local Unix" socket and a TCP socket. The network interface is run in the starting thread of the process and handles all incoming data asynchronously, dispatching complete requests to worker threads.

The network protocol (for both "Local Unix" and TCP connections) is binary, designed to minimize protocol parsing and the number of system calls necessary to process a request.

#### B. Worker threads

The daemon implements a pool of worker threads which accept requests from the network code, parse them and execute them. The threads communicate with the network interface using a protected (thread-safe) job queue. The number of threads is adjustable via command-line arguments. This feature includes a special support for "threadless" operation, in which the network interface calls the protocol parser as a function call in the same thread, eliminating synchronization overheads.

#### C. Data storage

Data storage is the most important part of the cache daemon as it has the biggest influence on its performance and capabilities. There are two large data structures implemented in the cache daemon.

The first is a hash table of static size whose elements (buckets) are roots of red-black trees which contain key-value pairs. These elements are protected by reader-writer locks (also called "shared-exclusive" locks). The hash table is populated by hashing the key portion of the pair. This mixing of data structures ensures a very high level of concurrency in accessing the data. Reader-writer locks per hash buckets allow for the highly desirable behaviour that readers (clients that only read data) never block other readers, greatly increasing performance for usual cache usage. The intention behind the design of this data structure was that, using a reasonably well distributed hash function, high concurrency of writers can also be achieved (up to the number of hash buckets). An illustration of the data storage organisation is presented in Fig. 1.

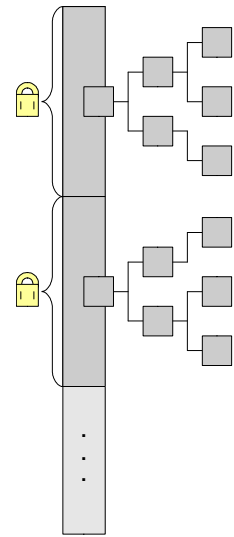

Fig. 1. Illustration of the key-value data structure used as the primary data storage pool

The second big data structure indexes the metadata tags for key-value records. It is a red-black tree of tag types (an integer value) whose elements are again red-black trees containing tag data (also an integer value) with pointers to the key-value records to which they are attached. The purpose of this organization is to enable performing queries on the data tags of the form "find all records of the given tag type" and "find all records of the given type and whose tag data conforms to a simple numerical comparison operation (lesser than, greater then)". The tree of data types is protected by a reader-writer lock and each of the tag data trees is protected by its own reader-writer lock, as illustrated in Fig. 2. Again, readers never block other readers and locking for write operations is localised in a way that allows concurrent access for tag queries.

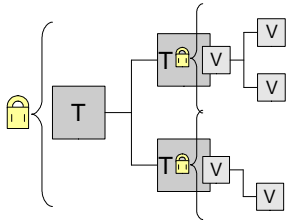

Fig. 2. Illustration of the tag tree structure

Records without metadata tags don't have any influence on or connection with the tag trees.

*D. Rationale and discussion*

We chose to implement both asynchronous network access and multithreading to achieve the maximum performance for the cache daemon [2]. This model is a hybrid of pure multi-process architecture (*MP*) and the even-driven architecture, and is sometimes called *asymmetric multi-process event-driven* (*AMPED*) [3]. In it, we dedicate a thread to network IO and accepting new connections. This model has been explored in part in [13], with the difference that our focus is on maximizing performed operations per second instead of network bandwidth. Our implementation tries hard to avoid unnecessary system calls, context switches and memory reallocation [11] [12]. The implementation has avoided most of protocol parsing overheads by using a binary protocol which includes data structure sizes and count fields in command packet headers.

Since the number of clients in the intended usage (web cache daemon) is relatively low (in the order of hundreds), we have avoided explicit connection scheduling described in [14]-[16].

We have opted for a thread-pool design (in which a fixed number of worker threads perform protocol parsing and data operations) to allow the administrator to tune the number of worker threads (via command line arguments) and thus the acceptable CPU load on the server. We have also implemented a special "threadless" mode in which there are no worker threads, but the network code makes a direct function call into the protocol parser, effectively making the daemon *single process event driven* (*SPED*). This mode can not make use of multiple CPUs, but is included for comparison with the other model.

As discussed in [7] and [9], the use of multi-processing and the relatively high standards we have set for concurrency of the requests have resulted in a need for careful choice of the structures and algorithms used for data storage. Traditional structures and algorithms used in caches, such as LRU and Splay trees [4], are not directly usable in high-concurrency environments. LRU and its derivatives need to maintain a global queue of objects, the maintenance of which needs to happen on every read access to a tracked object, which effectively serializes read operations. Splay trees radically change with every access and thus need to be exclusively locked for every access, serializing both readers and writers (much more seriously than LRU).

In order to maximize concurrency (minimize exclusive locking) and to limit the in-memory working set used during transactions (as discussed in [11]), we have chosen to use a combination of data structures, specifically a hash table and binary search trees, for the principal data storage structure. Each bucket of the hash table contains one binary search tree holding elements that hash to the bucket and a shared-exclusive locking object (*pthread rwlock*), thus setting a hard limit to the granularity of concurrency: write operations (and other operations requiring exclusive access to data) exclusively lock at most one bucket (one binary tree). Read operations acquire shared locks and do not block one another. The hash table is the principal source of writer concurrency. Given an uniform distribution of the hash function and significantly more hash buckets than there are worker threads (e.g. 256 vs. 4), the probability of threads blocking on data access is negligibly small, which is confirmed by our simulations. To increase overall performance and reduce the total number of locks, the size of the hash table is determined and fixed at program start and the table itself is not protected by locks. The garbage collector (which is implemented naively instead of a LRU-like mechanism) operates when the exclusive lock is already acquired (probabilistically, during write operations) and operates per hash-bucket. The consequence of operating per hash-bucket is a lower flexibility and accuracy in keeping track of the total size of allocated memory, and memory limits are forced to become per-bucket instead of per entire data pool.

The metadata tags structures design was driven by the same concerns, but also with the need to make certain query operations efficient (ranged comparison and grouping, i.e. less-than or greater-than). We have decided to allow the flexibility of queries on both the *type* and the *value* parts of metadata tags, and thus we implemented binary trees which are effective for this purpose.

## IV. SIMULATIONS

To aid in understanding of the performance and behaviour of the key-value store (the hash table containing binary search trees), we have created a GPSS simulation. The simulation models the behaviour of the system with a tunable number of worker threads and hash buckets. The simulated parts are: a task generator, worker threads, lock acquisition and release according to *pthread rwlock* semantics (with increased writer priority to avoid writer starvation) and the hash buckets. The task generator attempts to saturate the system. The timings used in the model are approximate and thus we're only interested in trends and proportions in the results. Fig. 3, 4 and 5 illustrate the percentage of "fast" lock acquisitions from the simulations, where "fast" is either uncontested lock acquisition or contested where total time spent waiting on a lock is deemed insignificant (less than 5% of average time spent holding the lock i.e. processing the task). Graphs are plotted with the number of hash buckets on the X-axis and a percentage of fast lock acquisitions on the Y-axis. Results cover the timings for both shared locks and exclusive locks, obtained during same simulations. The simulations were run for two cases: with 64 worker threads (which could today realistically be run on products such as the UltraSPARC T2 from Sun Microsystems [5]) and with 8 worker threads (which can be run on readily available servers on the common x86 platform [6] [10]). Individual figures describe the system behaviour with a varied ratio of reader and writer tasks.

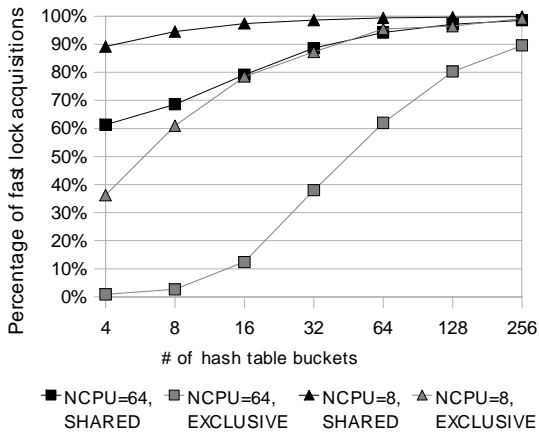

Fig. 3. Cache behaviour with 90% readers and 10% writers

Predictably, Fig 3. shows how a high ratio of hash buckets to threads makes almost all lock acquisitions fast.

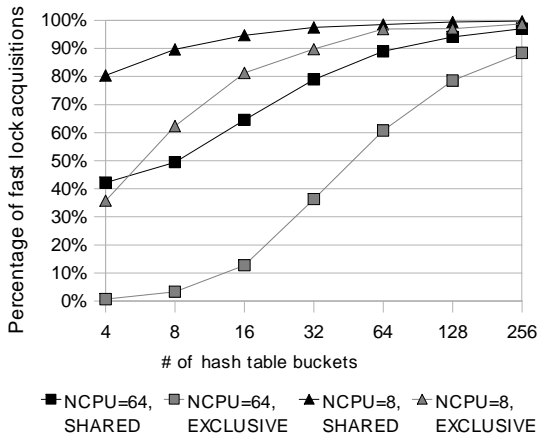

Fig. 4. Cache behaviour with 80% readers and 20% writers

Trends in the simulated system continue in Fig. 4 with expected effects of having a larger number of exclusive locks in the system. We observe that this load marks the boundary where having the same number of hash buckets and worker threads makes 90% of shared lock acquisitions fast.

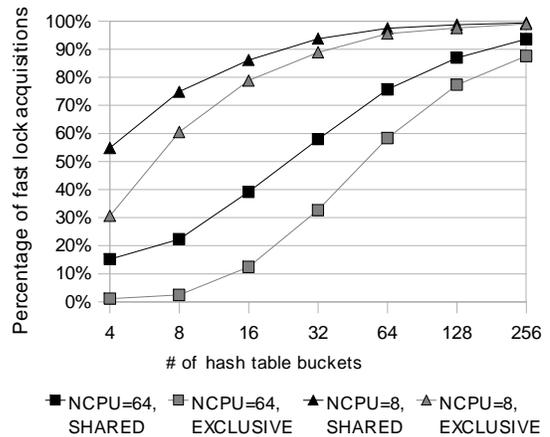

Fig. 5. Cache behaviour with 50% readers and 50% writers

In the situation presented in Fig. 5 the abundance of exclusive accesses (locks) in the system introduces significant increases in time spent waiting for hash bucket locks. Both kinds of locks are acquired with noticeable delays and the number of fast lock acquisitions falls appropriately.

The simulation results emphasise the lock contention, showing that equal relative performance can be achieved with the same ratio of worker threads and hash table buckets, and show an optimistic picture when the number of hash buckets is high. From these results, we have set the default number of buckets used by the program to 256, as that is clearly adequate for today's hardware. The graphs do not show the number of tasks dropped by the worker threads due to timeouts (simulated by the length of the queue in the task generator). Both types of simulated systems were subjected to the same load and the length of the task queue in systems with 8 worker threads was from 1.5 to 13 times as large as the same length in systems with 64 worker threads. This, coupled with simulated inefficiencies (additional delays) when lock contention is high between worker threads can have the effect of favouring systems with a lower number of worker threads (lock acquisition is faster because the contention is lower, but on the other hand less actual work is being done).

## V. Experimental Results

As this is a work in progress, we have performed only preliminary measurements of system performance and behaviour (of the key-value data store), on a limited variety of hardware.

To discover the impact of thread synchronization primitives, we benchmarked the program's performance on a single-CPU system with Pentium M @ 1.5 GHz, running FreeBSD 7.0, with both the daemon and the client on the same system, communicating via Unix Local sockets. As this is a single-CPU system, we present the results of measurements in "threadless" mode and with a single worker thread, to illustrate the tradeoffs present in the chosen architecture.

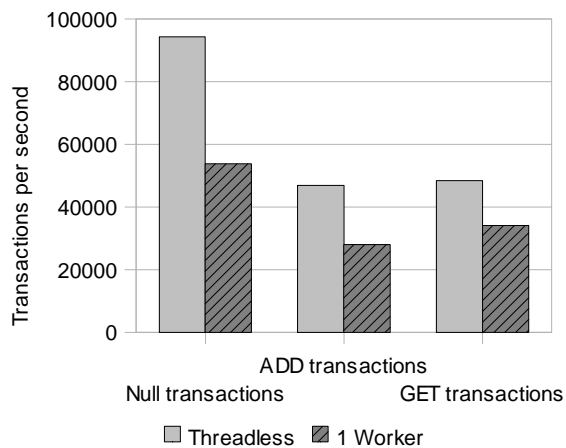

Fig. 6. Tradeoffs of multi-threading

Fig. 6 shows that the best option for single-CPU systems is the "threadless" mode (with a minimum 25% performance edge), in which the daemon degenerates into SPED-like behaviour. The costs of managing the thread-safe queue of tasks and the context switches involved in handing off the tasks from the network thread to the worker thread is high enough to result in noticeable slowdowns. The results lead us to conclude that in case of "Null transactions" (which are complete transactions, only without a payload command), these costs are almost the same as the processing time required for processing the transactions themselves.

Another type of benchmark was performed to explore the limits of performance of the cache daemon in its current implementation. These benchmarks use a mix of read and write operations (90% reads, 10% writes) on a precomputed data set of 30,000 records with size of 1 KB +/- 500 bytes, with a varied number of simultaneous clients.

TABLE I
BENCHMARK RESULTS OF THE MEMORY CACHE SERVER ON VARIOUS SYSTEMS AND LOADS

| System | No. clients | Ops / sec. |
|---|---|---|
| AMD Athlon 64 @ 1.8 GHz (32-bit), 2 core, FreeBSD 7.0, 2 worker threads, Local Sockets | 10 | 71,100 |
|  | 40 | 72,250 |
| Intel Core 2 Duo @ 1.8 GHz (32-bit), 2 core, Linux 2.6.22, 2 worker threads, Local Sockets | 10 | 75,000 |
|  | 40 | 79,700 |
| Two Intel Xeon 5320, 1.9 GHz (32-bit), 4 core systems, FreeBSD 7.0, 4 worker threads, Remote TCP (gigabit Ethernet) | 10 | 95,150 |
|  | 40 | 113,650 |

The results presented in Table 1 are promising and adequate for many "real-world" purposes, however we believe that there is room for improvement and that testing on faster hardware with more CPU cores may yield information about possible areas of improvement in performance and scalability (which is on our future research agenda).

We have performed a preliminary comparison of our memory cache server to an existing solution, Memcached 1.2.1, used by many existing high-performance web sites [1] [8], with the same data set as used for results in Table 1 and on the system from the first row in the table.

TABLE II
BENCHMARK RESULTS OF OUR MEMORY CACHE SERVER COMPARED TO MEMCACHED

|  | No. Clients | Ops / sec. |
|---|---|---|
| Our cache server, 2 worker threads | 10 | 71,100 |
| Memcached, threadless | 10 | 35,150 |

We attribute the differences in performance presented in Table 2 to the inefficient text network protocol used by Memcached and a design that doesn't scale well to multi-CPU systems.

VI. CONCLUSION

This paper presents the design and implementation of a high-performance memory cache database server. In its creation we have designed many optimizations, including data structures permitting highly concurrent operations, multi-threaded core based on the thread-pool model and an optimized network communication model. We have analysed and simulated the designed structures and algorithms, adapted and implemented it, and performed benchmarks of the resulting server program.

The intended usage for this server is as an external cache database for web applications, and preliminary analysis of its performance and behaviour suggests that the current implementation of the server is sufficient for this purpose.

The result of this project is a directly usable product which will soon be implemented in our Faculty's web applications.

## VIII. BIOGRAPHIES

**I. Voras** (M'06), was born in Slavonski Brod, Croatia. He received Dipl.ing. in Computer Engineering (2006) from the Faculty of Electrical Engineering and Computing (FER) at the University of Zagreb, Croatia. Since 2006 he has been employed by the Faculty as an Internet Services Architect and is a graduate student (PhD) at the same Faculty, where he has participated in research projects at the Department of Control and Computer Engineering. His current research interests are in the fields of distributed systems and network communications, with a special interest in performance optimizations. He is an active member of several Open source projects and is a regular contributor to the FreeBSD operating system. Contact e-mail address: ivan.voras@fer.hr.

**D. Basch** was born in Rijeka, Croatia. He received Dipl.ing. in Electrical Engineering (1991), M.Sc. (1994) and Ph.D. (2000) in Computer Science from the Faculty of Electrical Engineering and Computing (FER) at the University of Zagreb (Croatia). In 1992 he joined Department of Control and Computer Engineering (at FER) as a researcher. At present he works at the same department as an associate professor. His research interests include programming language design and implementation, garbage collection algorithms, and modelling and simulation. Contact e-mail address: danko.basch@fer.hr.

**M. Žagar** (M'93-SM'04), professor of computing at the University of Zagreb, Croatia, received Dipl.ing., M.Sc.CS and Ph.D.CS degrees, all from the University of Zagreb, Faculty of Electrical Engineering and Computing (FER) in 1975, 1978, 1985 respectively. In 1977 M. Žagar joined FER and since then has been involved in different scientific projects and educational activities.

He received British Council fellowship (UMIST - Manchester, 1983) and Fulbright fellowship (UCSB - Santa Barbara, 1983/84). His current professional interests include: computer architectures, design automation, real-time microcomputers, distributed measurements/control, ubiquitous/ pervasive computing, open computing (JavaWorld, XML,..).

M. Žagar is author/co-author of 5 books and about 100 scientific/ professional journal and conference papers. He is senior member in Croatian Academy of Engineering. In 2006 he received "Best educator" award from the IEEE/CS Croatia Section.. Contact e-mail address: mario.zagar@fer.hr.